\documentclass[prl,twocolumn,superscriptaddress,showpacs,amsmath,amssymb,amsfonts]{revtex4}

\usepackage[sort&compress]{natbib}

\begin{document}

\title{A Generalized No-Broadcasting Theorem}

\author{Howard Barnum} \email{barnum@lanl.gov} \affiliation{CCS-3:
Information Sciences,
MS B256, Los Alamos
National Laboratory, Los Alamos, NM 87545 USA}

\author{Jonathan Barrett} \email{jbarrett@perimeterinstitute.ca}
\affiliation{Perimeter Institute for Theoretical Physics, 31 Caroline
Street N, Waterloo, Ontario N2L 2Y5, Canada}

\author{Matthew Leifer} \email{matt@mattleifer.info}
\affiliation{Perimeter Institute for Theoretical Physics, 31 Caroline
Street N, Waterloo, Ontario N2L 2Y5, Canada}
\affiliation{Institute for Quantum Computing, University of Waterloo, 200 University Ave. W., Waterloo, Ontario N2L 3G1, Canada} 

\author{Alexander Wilce} \email{wilce@susqu.edu} \affiliation{Department of
Mathematical Sciences, Susquehanna University, Selinsgrove, PA 17870
USA}

\begin{abstract}
We prove a generalized version of the no-broadcasting theorem,
applicable to essentially \emph{any} nonclassical finite-dimensional
probabilistic model satisfying a no-signaling criterion, including ones
with ``super-quantum'' correlations.  A strengthened
version of the quantum no-broadcasting theorem follows, and its proof is
significantly simpler than existing proofs of the
no-broadcasting theorem.
\end{abstract}

\maketitle

The no-cloning theorem \cite{Wootters:1982}, as sharpened in \cite{Yuen:1986}, 
states that there is no way of 
blindly copying a pair of nonorthogonal pure states.  
More precisely, for any pair of
nonorthogonal pure states $\rho_i, i \in \{1,2\}$, there is no
trace-preserving completely positive map $\mathcal{E}$ such that
$\forall i, \mathcal{E}(\rho_i)=\rho_i\otimes\rho_i$.  In a classical
context, with probability distributions replacing density operators,
universal cloning of pure states is possible (though in the
infinite-dimensional case, this may be an unphysical idealization
\cite{DPP02,WalkerBraunstein06}), but even here, universal cloning
producing {\em independent} copies is impossible if we require mixed
states to be cloned also.  A set of mixed states that {\em can} be
cloned, in classical or quantum mechanics, must be mutually
orthogonal---their density matrices or probability distributions must
have nonoverlapping support \cite{Barnum:1996}.

Broadcasting is a weaker notion than cloning.   A map
$\mathcal{E}$ that takes states on $\mathcal{H}$ to states on
$\mathcal{H}_A\otimes \mathcal{H}_B$
\emph{broadcasts} a state $\rho$ if $\mathrm{Tr}_A
\left(\mathcal{E}(\rho)\right)=\mathrm{Tr}_B
\left(\mathcal{E}(\rho)\right)= \rho$, i.e. we do not require that the final
state is a product state. A set of states is
\emph{broadcastable} if and only if there is a map $\mathcal{E}$ that
broadcasts each state in the set
\footnote{While the $N$-system to $M>N$-system generalization of perfect
cloning is impossible in the quantum case, a 
similar generalization of broadcasting, taking states with
$N>1$ identical marginals to ones with $M$ of the same marginals {\em is}
sometimes possible \cite{D'Ariano:2005}.
}.
This generalizes pure-state cloning to mixed states in a way that 
picks out ``classical'' sets of states 
better than requiring independent copies does:
the no-broadcasting
theorem \cite{Barnum:1996} says not only that universal
broadcasting is impossible, but also that a
set of states is broadcastable if and only if they commute pairwise.  For the analogous classical notions, universal broadcasting
can be achieved (with minor 
caveats about the infinite dimensional
case \cite{WalkerBraunstein06}).

Cloning and broadcasting are two of the most elementary
information-theoretic tasks, and the impossibility of universal
cloning and broadcasting are two of the more significant ways in which
quantum theory differs from classical theory.  In order better to 
understand these differences, and how they may underpin the better-than-classical
performance of quantum information processing, it is helpful to consider
the information-theoretic properties of probabilistic theories that are neither
classical nor quantum.
This has lately been a very active area of investigation in quantum
information theory \cite{Hardy2001a, Barrett:2005a, Barrett:2005,
BBLMTU05, LPSW06}.

Therefore, it is natural to ask whether the no-broadcasting theorem is
a special feature of quantum theory, i.e. something that identifies it
uniquely from a space of surrounding alternative theories, or generic,
so that the \emph{possibility} of universal broadcasting is special to
\emph{classical} theories.  We show that the latter is the case.
Working in a very general framework that assumes little apart from the
convexity and finite dimension of state spaces, and the no-signaling
principle for bipartite systems, we show that a set of states is
broadcastable if and only if it is contained in a simplex generated by
states that are jointly distinguishable via a single-shot measurement.
This reduces to the standard no-broadcasting theorem in the quantum
case, and implies that universal broadcasting is only possible in
classical theories.  The resulting proof of the quantum
no-broadcasting theorem is significantly simpler than the original
proof of \cite{Barnum:1996}, and more intuitive and
self-contained than that based on Lindblad's theorem
\cite{Lindblad:1999} (although the latter provided useful ideas).

\emph{An operational framework.} Before proving the main theorem, we
describe a mathematical formalism involving only minimal constraints
that accommodates a wide range of possible probabilistic theories.
Our approach is based on convex sets, and has a long pedigree
\cite{Mackey:1963, Foulis:1981, Hardy2001a}, although the
investigation of information processing in this context is more recent
(see \cite{Barnum2002a} and in particular \cite{Barrett:2005}, where a
no-cloning theorem is derived for all nonclassical theories).  The
framework is broad enough to include theories with
``Popescu-Rohrlich'' or ``nonlocal boxes'' \cite{Barrett:2005a,
BBLMTU05}, exhibiting stronger-than-quantum correlations.

The basic approach is operational, meaning that notions such as
\emph{system}, \emph{state}, and \emph{measurement} are among 
its fundamental concepts.  We associate with a given type of system a set
$\Omega$ of possible states. If it is
possible to prepare a system in a state $\omega_1$ or $\omega_2$, then it should be possible to
prepare any probabilistic mixture of the two (say by tossing a biased
coin, preparing one or the other according to the outcome, and then
forgetting the outcome) so we assume that $\Omega$ is convex.

Whatever else a state is, it should define probabilities for
measurement outcomes.   We write the probability of getting outcome $e$ when
the state is $\omega$ as $e(\omega)$. Suppose that $\omega=p\,\omega_1
+ (1-p)\,\omega_2$ and that this mixed state is prepared by tossing a
biased coin, as above. The probability of outcome $e$
should be the weighted average of the probabilities of $e$
given $\omega_1$ and $\omega_2$, i.e., $e(\omega)=p\,
e(\omega_1)+(1-p)\, e(\omega_2)$. Therefore we identify $e$ with an affine
linear functional $\Omega\rightarrow [0,1]$. We refer to any such
functional as an \emph{effect}. The \emph{unit effect} $u$ is defined
by $u(\omega)=1$ for all $\omega\in\Omega$, and represents a
measurement outcome that is certain to occur no matter what the
state.  The set of all effects is denoted $[0,u]$.
A measurement corresponds to
a set of effects $\{e_i\}$ such that $\sum_i e_i = u$, that is $\sum_i
e_i(\omega)=1$ for all $\omega\in\Omega$.

In quantum theory, $\Omega$ is the set of density operators on some
Hilbert space. A particular measurement outcome is associated with a
positive operator $E$ bounded by $0$ and the identity $I$, such that
the probability of getting this outcome for state $\rho$ is given by 
$\mathrm{Tr}(E\,\rho)$.  This does define an affine linear functional
on the state space since if $\rho=p\, \rho_1 + (1-p)\, \rho_2$, then
$\mathrm{Tr}(E\,\rho)=p\,\mathrm{Tr}(E\,\rho_1)+(1-p)\,\mathrm{Tr}(E\,\rho_2)$. The
unit effect corresponds to $I$ and a measurement as a whole 
corresponds to a set
of positive operators summing to $I$, i.e. a discrete POVM.

A transformation in quantum theory corresponds to a linear
trace-preserving completely positive map
taking states on a
Hilbert space $\mathcal{H}$ into states on a Hilbert space
$\mathcal{H}'$ \footnote{We ignore
probabilistic transformations, which correspond to trace-decreasing
completely positive maps}.  
As with the rule for measurement outcomes, linearity
ensures that transformations respect probabilistic mixtures of
states. In the generalized framework, a transformation corresponds
to an affine mapping $T:\Omega\rightarrow\Omega'$, where $\Omega$ is
the state space of the system prior to the transformation, 
and $\Omega'$ is the post-transformation state space. One
should not assume that {\em all} such affine maps correspond to 
allowed transformations in a particular theory. For example, in
quantum theory only completely positive maps (not arbitrary
positive maps) do.

The set of all affine
functionals $\Omega\rightarrow {\Bbb R}$ is a vector space denoted
$A(\Omega)$. There is a natural embedding of $\Omega$ in
$A(\Omega)^\ast$ (the dual space of $A(\Omega)$), given by
$\omega \mapsto \hat{\omega}$, where $\hat{\omega}(a) = a(\omega)$ for
all $a \in A(\Omega)$. This enables us to identify $\omega$ with
$\hat{\omega}$, writing either $\omega(a)$ or $a(\omega)$ as
convenient. Let $V(\Omega)$ be the linear span of $\Omega$ in
$A(\Omega)^\ast$. Then, $\Omega$ is finite-dimensional iff
$V(\Omega)$ is finite-dimensional. 
We assume state spaces are
finite-dimensional and compact, which guarantees that $\Omega$ is the closed convex
hull of its extreme points (referred to as \emph{pure}
states).

A $d$-dimensional system is \emph{classical} iff $\Omega$ is the
convex hull of $d+1$ linearly independent pure states (a simplex), in
which case $\Omega$ can be thought of as the set of probability
distributions over $d+1$ distinct possibilities.  Only in such systems can the
extremal points be perfectly distinguished from each other by a single
measurement, a point discussed in the proof of Theorem 2 below.
Classical systems are also characterized by the fact that each state
has a unique decomposition into extremal states.  A theory is 
classical iff each system in the theory is classical.

\emph{Joint systems.} Suppose systems $A$ and $B$ have
state spaces $\Omega_A$ and $\Omega_B$. The joint system $AB$ will
have its own state space, $\Omega_{AB}$, but how are $\Omega_A$,
$\Omega_B$ and $\Omega_{AB}$ related? Assume: (i) a joint state
defines a joint probability for each pair of effects $(e_A,e_B)$,
where $e_A\in A(\Omega_A)$ and $e_B\in A(\Omega_B)$ \footnote{If a
theory specifies a set of ``physically allowed'' measurements, with
outcomes corresponding to a proper subset of the effects, then one
might weaken assumption (i) to require only that joint states assign
probabilities to pairs of effects representing physical measurement
outcomes. Whether our no-broadcasting theorem would still hold for
such theories, we leave open.}; (ii) these joint probabilities respect
the no-signaling principle, i.e., the marginal probabilities for the
outcomes of a measurement on $B$ do not depend on which measurement
was performed on $A$ and vice versa; (iii) if the joint probabilities
for all pairs of effects $(e_A,e_B)$ are specified, then the joint
state is specified.

These assumptions do not determine $\Omega_{AB}$ uniquely in general,
but they do imply \cite{Foulis:1981, Klay:1987, Wilce:1992,
Barrett:2005} that it must be a convex set whose span can be
identified with the vector space $V(\Omega_A)\otimes
V(\Omega_B)$. Further, it must lie between two extremes, the
\emph{maximal} and the \emph{minimal tensor product}.
The maximal tensor product, $\Omega_A \otimes_{max} \Omega_B$, is the
set of all bilinear functionals $\phi:A(\Omega_A)\times
A(\Omega_B)\rightarrow {\Bbb R}$ such that (i) $\phi(e,f)\geq 0$ for
all pairs of effects $(e,f)$, and (ii) $\phi(u_A,u_B)=1$, where $u_A$
and $u_B$ are the unit effects for systems $A$ and $B$. 
The maximal tensor product has an important operational
characterization: it is the largest set of states in $(A(\Omega_A) \otimes 
A(\Omega_B))^*$ assigning
probabilities to all product measurements but not allowing signaling
\cite{Foulis:1981,Klay:1987}.
The minimal
tensor product, $\Omega_A \otimes_{min} \Omega_B$, is the convex hull
of product states, where a product $\omega_A\otimes\omega_B$ is
defined by $(\omega_A\otimes\omega_B) (a,b) = \omega_A(a)\omega_B(b)$
for all pairs $(a,b)\in A(\Omega_A)\times A(\Omega_B)$.  These 
appeared in \cite{Namioka:1969} in the context of abstract compact
convex sets.  

Joint states in
$\Omega_A\otimes_{min}\Omega_B$ are \emph{separable} and those in
$\Omega_A\otimes_{max}\Omega_B$ but not in
$\Omega_A\otimes_{min}\Omega_B$ are \emph{entangled}.

A particular theory should specify, besides $\Omega_A$ and $\Omega_B$,
a set of joint states $\Omega_{AB}$ such that
$\Omega_A\otimes_{min}\Omega_B\subseteq
\Omega_{AB}\subseteq\Omega_A\otimes_{max}\Omega_B$.  We call this
$\Omega_A\otimes\Omega_B$, keeping in mind that this may be any convex
set bounded by the minimal and maximal tensor products. If either $A$
or $B$ is classical then $\Omega_A\otimes_{min}\Omega_B =
\Omega_A\otimes_{max}\Omega_B$ and there is no entanglement.  In
particular, if both are classical then both $\Omega_A \otimes_{min}
\Omega_B$ and $\Omega_A \otimes_{max} \Omega_B$ are the simplex whose
vertices are ordered pairs of an extremal point of $\Omega_A$ and
one of  $\Omega_B$.  For quantum theory
$\Omega_A\otimes_{min}\Omega_B\subset \Omega_{AB}\subset
\Omega_A\otimes_{max}\Omega_B$, where the inclusions are strict
\cite{Foulis:1981,Klay:1987,Klay:1988,Wilce:1992,Barnum:2005}.


This treatment is easy to generalize to multipartite
systems, by allowing $A$ and $B$ themselves to be
composite.   Also, we can define unambiguously the notion of a reduced
(or marginal) state. Any state $\omega_{AB}\in
\Omega_A\otimes_{max}\Omega_B$ has reduced states $\omega_A$ and
$\omega_B$ defined such that $\omega_A(a) = \omega_{AB}(a,u_B)$ and
$\omega_B(b) = \omega_{AB}(u_A,b)$. It is easy to show that if either
reduced state is pure, then $\omega_{AB}=\omega_A\otimes\omega_B$.

\emph{Cloning and Broadcasting.}  We can generalize the definitions of
cloning and broadcasting given at the beginning.  Consider a state
space $\Omega$ and a transformation $T:\Omega\rightarrow\Omega\otimes
\Omega$. Denote the reduced states of $T(\omega)$ by $(T(\omega))_A$
and $(T(\omega))_B$.  We say $T$ \emph{clones} a state $\omega$ iff
$T(\omega)=\omega\otimes \omega$. A set of states is \emph{cloneable}
iff there is a single map $T$ such that $T$ clones each state in the
set.  We say $T$ \emph{broadcasts} a state $\omega$ iff
$(T(\omega))_A=(T(\omega))_B=\omega$.  And a set of states is
\emph{broadcastable} iff there is a single map $T$ such that $T$
broadcasts each state in the set.  In addition, we say a set of states
$\{\omega_1,\ldots,\omega_n\}$ is \emph{jointly distinguishable} iff
they can be distinguished with certainty with a single-shot
measurement, i.e., there exists a measurement $E$ with outcomes
$e_1,\ldots,e_n$ such that
$\omega_i(e_j)=\delta_{ij}$.

\emph{Theorem 1.} For any finite dimensional state space $\Omega$ and
any choice of tensor product $\Omega\otimes \Omega$, a set of states
is cloneable iff it is jointly distinguishable.

A rigorous proof of this theorem is given in \cite{Barnum:2006}.
Intuitively, if a set of states is cloneable,
then they may be distinguished by making many clones 
and then 
identifying the state with suitable measurements on the
copies. 
Conversely, if the states are jointly distinguishable,
then one way of cloning is to perform the measurement that distinguishes them
and then to prepare two copies.

\emph{Theorem 2.}
Universal cloning is only possible for classical systems.

\emph{Proof.} Universal cloning implies that the set of all pure
states is clonable, so any finite subset
$\{\omega_1,\ldots,\omega_n\}$ is clonable.  From Theorem~1, it
follows that this set is jointly distinguishable, thus we can find
affine functionals $e_1,\ldots,e_n$ such that
$\omega_i(e_j)=\delta_{ij}$. It follows that the $\omega_i$ 
are linearly independent in 
$V(\Omega)$. Since this holds for any finite subset of pure states, all
pure states are linearly independent, and since $V(\Omega)$ is finite
dimensional there can only be a finite number of such states. Thus
$\Omega$ is a simplex and the system is classical. $\Box$

For any kind of system, broadcasting of pure states reduces
to cloning because if $\omega$ is pure
and $T(\omega)$ has reduced states equal to $\omega$, then $T(\omega)
= \omega \otimes \omega$.   So 
Theorem~2 implies that universal broadcasting is possible only for 
classical systems.  Our main theorem goes further: it
specifies exactly when a set of states is broadcastable.

\emph{Theorem 3.}
A set of states is broadcastable iff it lies in a simplex
generated by jointly distinguishable states. 


\emph{Proof.} 
For the ``if'' direction, it is easily verified that the map that
clones the extreme points of the simplex (cf. the discussion following Theorem 1)
broadcasts the entire simplex.   
For ``only if'', consider 
$B':\Omega\rightarrow\Omega\otimes\Omega$, and denote by $\Gamma'$ the set
of states broadcast by $B'$.  It is immediate from the definition of
broadcasting that if $B'$ broadcasts both $\omega_1$ and
$\omega_2$, then $B'$ broadcasts any convex combination of them. Thus
$\Gamma'$ is convex. Let $\sigma : \Omega\otimes\Omega \rightarrow
\Omega\otimes\Omega$ be the swap operation, defined by
$\sigma(\omega_A\otimes \omega_B)=\omega_B\otimes\omega_A$. Define the
symmetrized map $B : \Omega \rightarrow \Omega \otimes \Omega$ by $B = (B'
+ \sigma \circ B')/2$. Denote by $\Gamma$ the set of states broadcast
by $B$ and note that $\Gamma' \subseteq \Gamma$. The strategy will be
to define, in terms of $B$, a map
$Q:\Gamma\rightarrow\Gamma\otimes_{max}\Gamma$ such that $Q$ is
universally broadcasting on $\Gamma$, hence cloning for all states
extremal in $\Gamma$.  It is critical that $Q$'s 
domain be $\Gamma$, rather than $\Omega$, 
because the extremal points of $\Gamma$ are not necessarily 
extremal in $\Omega$, and we need to use Theorem 1 to conclude that $\Gamma$'s 
extremal points are cloned.

\emph{Definition.}  A {\em compression} of $\Omega$ onto a subset $\Gamma$ is an
idempotent affine mapping $\Omega \rightarrow \Omega$ having range
$\Gamma$.

\emph{Lemma 1.}
Let $T : \Omega \rightarrow \Omega$
be any transformation taking $\Omega$ into itself. Then there exists a
compression of $\Omega$ onto the set of fixed points of $T$.
%

\emph{Proof.} For each $n \in {\Bbb N}$, let $P_{n} = \frac{1}{n} \sum_{k=1}^{n}
T^{k} : \Omega \rightarrow \Omega$. Since $\Omega$ is compact, we may
assume (passing to a subsequence if necessary) that $(P_{n})$
converges to a limiting affine map $P : \Omega \rightarrow \Omega$.  If
$T(\omega) = \omega$, then clearly $P(\omega) = \omega$; conversely,
if $\omega = P(\phi)$ for some $\phi \in \Omega$, then 
$T(\omega) = \lim_{n \rightarrow \infty} \frac{1}{n} \sum_{k=1}^{n}
T^{k+1}(\phi) = \lim_{n \rightarrow \infty} \frac{1}{n} \sum_{k=1}^{n+1}
T^{k} (\phi)- \lim_{n \rightarrow \infty} \frac{1}{n} T(\phi)$.  
We note that $\lim_{n \rightarrow \infty} \frac{1}{n} T(\phi)=0$, and 
rewrite the first term as 
$\lim_{n \rightarrow \infty} \frac{1}{n} \sum_{k=1}^n T^k (\phi)
+ \lim_{n \rightarrow \infty}
\frac{1}{n} T^{n+1}(\phi) = P(\phi) = \omega$.
Thus, the range of $P$ is exactly the fixed-point set of $T$, as
claimed.  Therefore, as $P(\phi)$ is a fixed point of $T$, we
have $P(P(\phi)) = P(\phi)$ for any $\phi$, i.e., $P$ is
idempotent. $\Box$

Continuing with the proof of Theorem 3, 
recall the symmetrized map $B$, which broadcasts states in $\Gamma$,
and define a map $B_A$ so that $B_A(\omega)$ is given by the reduced
state $(B(\omega))_A$. Note that $\omega \in \Gamma$ iff $\omega$ is a
fixed point of $B_A$. By Lemma~1, there is a compression 
$P$ onto $\Gamma$. Consider $P$ as a
map $\Omega\rightarrow\Gamma$; there is a unique definition of
$P\otimes P:\Omega\otimes_{max}\Omega\rightarrow\Gamma\otimes_{max}\Gamma$
satisfying $(P\otimes P)(\omega_A\otimes\omega_B)=P(\omega_A)\otimes
P(\omega_B)$. Now define $Q:\Gamma\rightarrow \Gamma\otimes_{max}\Gamma$
by $Q(\gamma) = (P \otimes P)(B(\gamma))$.
$Q$ is universally broadcasting on $\Gamma$. For
if $\gamma \in \Gamma$, we have, for all $e_{\Gamma} \in [0,u_{\Gamma}]$,
\begin{align*}
& Q_A (\gamma)(e_{\Gamma}) = Q(\gamma)(e_{\Gamma} \otimes u_{\Gamma})
= ((P \otimes P) B(\gamma))(e_{\Gamma} \otimes u_{\Gamma}) \\ & =
B(\gamma)(P^{\ast} e_{\Gamma} \otimes P^{\ast} u_{\Gamma} ) =
B_A(\gamma)(P^{\ast} e_{\Gamma}) = \gamma(P^{\ast} e_{\Gamma}) \\ & =
P(\gamma)(e_{\Gamma}) = \gamma(e_{\Gamma}),
\end{align*} 
where $P^{\ast} e:= e\circ P$ for arbitrary $e$, and the last step
uses the fact that $P(\gamma) = \gamma$, since $\gamma \in \Gamma$. 
(Here $[0, u_\Gamma]$ is defined analogously to $[0, u]$, but with $\Gamma$ playing
the role played by $\Omega$.)
It follows that $Q_A(\gamma) = \gamma$; similarly, 
$Q_B(\gamma) = \gamma$.
Since $Q$ is universally broadcasting on
$\Gamma$, it broadcasts $\Gamma$'s extremal states. Broadcasting
reduces to cloning for extremal states, so $Q$ is
universally cloning on the set of extremal points of 
$\Gamma$. 
It follows from
Theorem~1 that the extreme points of $\Gamma$ are
jointly distinguishable in $\Gamma$, i.e. via an observable consisting of
effects in $[0, u_\Gamma]$.  We can ``lift'' effects 
$e_\Gamma \in [0,u_\Gamma]$, defined only on the
span $V(\Gamma)$ of $\Gamma$, to functionals on $V(\Omega)$ via the
map $e_\Gamma \mapsto e_\Gamma \circ P$, which takes effects in
$[0,u_\Gamma]$ to ones in $[0, u]$.  Thus any observable 
on $\Gamma$ lifts via $P$ to one on $\Omega$.  
This includes the one that 
distinguishes the extreme points of $\Gamma$, so Theorem~3 follows. 
$\Box$

The proof uses the symmetrized $B$, rather than $B'$, 
because the set $\Gamma$ of states broadcast by $B$ is the fixed-point set of
$B$'s marginal maps $B_A = B_B$, so Lemma 1 provides a  
compression $P$ onto $\Gamma$.
The set $\Gamma'$ broadcast by $B'$ is the
{\em intersection} of the fixed-point sets of the marginal maps
$B'_A$ and $B'_B$ which need not be equal, so Lemma 1 does not 
provide a compression onto it. 

\emph{Corollary (quantum no-broadcasting theorem).}  
Let $\Gamma$ be a set of density operators on a Hilbert space
$\mathcal{H}$.  If there is a positive map $T : {\cal
B}(\mathcal{H}) \rightarrow {\cal B}(\mathcal{H})$ broadcasting each $\rho
\in \Gamma$ then the operators in $\Gamma$ mutually commute.

\emph{Proof.} By Theorem~3, $\Gamma$ is
contained in a simplex generated by distinguishable, hence commuting,
density operators. Hence the operators in $\Gamma$ also
commute. $\Box$

This result is stronger than that in \cite{Barnum:1996}, which applied to
\emph{completely positive} maps rather than all positive maps.

Theorem~3 tells us little about the convex structure of the set
$\Gamma$ of states broadcast by a map $B$.  But
Ref.~\cite{Barnum:2006} builds on it to show that any such $\Gamma$
\emph{is} a simplex generated by jointly distinguishable states.

\emph{Conclusion.}  In order to understand the nature of information
processing in quantum mechanics, it is useful to demarcate those
phenomena that are {\em essentially} quantum, from those that are more
generically non-classical.  This Letter has identified an important
feature of quantum information that is generic: the no-broadcasting
theorem.  Note that not {\em every} qualitative result of quantum
information is similarly generic, e.g. teleportation is not possible
in every probabilistic theory \cite{Barrett:2005, Short:2005a}.


In \cite{Clifton:2003} it was shown that the conjunction of
no-signaling, no-broadcasting and no-bit-commitment implies the
existence of noncommuting observables and entangled states for
theories in a $C^*$-algebraic framework, yielding theories quite close
to quantum theory.  However, this framework is already close to
quantum theory, since all theories in it have Hilbert space
representations and the finite-dimensional ones are just quantum
theory, classical probability and quantum theory with superselection
rules.  The framework adopted in this Letter is more natural for
pursuing the program of deriving quantum theory from information
theoretic axioms \cite{Fuchs:2002, Clifton:2003}, as it is narrow
enough to allow axioms to be succinctly expressed mathematically, but
broad enough that the main substantive assumptions will be contained in the
axioms rather than in the framework itself.  The framework
\emph{assumes} no-signaling, and we have shown that no-broadcasting
holds for any nonclassical model within it.  Such models can be very
different from quantum theory, e.g. they may support stronger-than-quantum
correlations \cite{Barrett:2005a}.  An open question is whether
no-bit-commitment is also generic, but in any case it seems unlikely
that these three axioms alone would get one particularly close to
quantum theory.  Thus our results suggest that future progress in
characterizing quantum theory in terms of information-theoretic tasks
is likely to require assumptions of a less generic character, such as
the possibility of teleportation.

\noindent{\bf Acknowledgments:} Research at 
Perimeter Institute for Theoretical Physics
is supported 
in part 
by the Government of
Canada through NSERC and 
by 
the Province of Ontario through MRI.
At IQC, ML is supported
by MITACS and ORDCF.  Part of this work was completed while ML was at CQC, University of 
Cambridge, supported by the EC through QAP IST-3-015848, and through the FP6-FET
project SCALA, CT-015714.

\bibliographystyle{apsrev} 

\end{document}